\documentclass[12pt]{article}

\begin{document}

%\hspace{8cm}odnor-ur.tex

\title{Canonical Quantization\\ of the tachyon field}
\author{Efimov G.V. \vspace*{0.2\baselineskip}\\
 \itshape LTPh, JINR,
141980 Dubna, Russia\vspace*{0.2\baselineskip} }
%
% \date{Pacs Numbers: 13.20.-v, 13.20.Fc, 13.20.He, 24.85.+p}
%
\maketitle
%..........

\begin{abstract}
The canonical quantization of the tachyon field is suggested.
Quantization is based on the conception of stable and unstable
components of the tachyon degrees of freedom.
\end{abstract}

\section{Introduction}

Last decade one can see a revival of interest to particles moving
with faster-than-light speed with spectrum energy
$\epsilon_k=\sqrt{k^2-m^2}>0$ (see \cite{Kos, Chi}). This particle
is named a tachyon.

A classical tachyon arises in inflationary models based on the
string theory and plays a definite role to explain the physics of
black holes. However, recent experiments indicate the possibility
that neutrino moves with faster-than-light speed. So we need a
reasonable description of the tachyon field in classical and quantum
physics. Usually, a tachyon is connected with the relativistic
Klein-Gordon equation with negative square mass $m^2<0$ in order to
guarantee faster-than-light speed. There are some suggestions how to
quantize the tachyon field (see, for example, \cite{Rem}).

In this paper, we formulate a new variant of quantization.

What is the quantization in the standard quantum field theory? The
main and, unfortunately, unique practical aim is to construct the
scattering $S$-matrix which describes all possible transitions from
some free states to other free states of particles. For this aim, we
need to quantize free particle fields and calculate so-called
propagator, or causal Green functions of fields under consideration.
Then using  a given interaction Lagrangian one can construct the
$S$-matrix in the form of perturbation expansion, each term of which
is represented as a set of appropriate Feynman diagrams. A diagram
is a product of appropriate propagators. The next problem is to
prove that this construction satisfies all necessary conditions:
relativistic covariance, unitary, causalitity and so on. We will
follow this scheme and our aim is to quantize the free tachyon field
defined on an appropriate Fock space and calculate the tachyon
causal Green function.

Our idea is quite simple. Any usual free scalar field is a set of
oscillators with real frequencies $\omega_k^2=k^2+m^2$ for all real
$k$, so that the scalar field quantization is a generalization of
the standard quantization of oscillators  to the system with an
infinite number of degrees of freedom. In the case with $m^2<0$ the
tachyon field consists of two parts, one of which contains
oscillators with real frequencies $\omega_k^2=k^2-m^2>0$ and the
second contains ''unstable'' oscillators with imaginary frequencies
$\omega_k^2=-(m^2-k^2)<0$. Thus, the problem is how to perform the
quantization of this ''unstable'' part.

In this paper, we will show how to quantize the ''unstable''
oscillators with $\omega^2<0$ and apply this procedure to the
tachyon field.

\section{Classical oscillator}

First of all, let us consider the classical stable and unstable
motion. We have two Lagrangians
\begin{eqnarray}\label{cl1}
&& L_{st}={\dot{q}^2(t)\over2}-{\omega^2q^2\over2},~~~~
L_{un}={\dot{q}^2(t)\over2}+{\Omega^2q^2\over2}.
\end{eqnarray}
The Lagrangian $L_{st}$ describes the stable finite motion and
$L_{un}$ corresponds to the unstable unbounded motion. The equations
are
\begin{eqnarray}\label{cl1}
&& \ddot{q}_{st}(t)+\omega^2q_{st}(t)=0,~~~~
\ddot{q}_{st}(t)-\Omega^2q_{st}(t)=0.
\end{eqnarray}
The solutions for $q_{st}(t)$ and $q_{un}(t)$ look like
\begin{eqnarray}\label{cl1}
&& q_{st}(t)=D_1\cos(\omega t)+D_2\sin(\omega t),\\
&& q_{un}(t)=C_1e^{-\Omega t}+С_2e^{\Omega t}.\nonumber
\end{eqnarray}
Both solutions for $q_{st}(t)$ are bounded and one can formulate the
Cauchy problem. However, in the second case for $q_{un}(t)$ we have
increasing  and decreasing, or damped, solutions. One can formulate
the problem as follows: we look for the damped solutions only. It
means that we should put $C_2=0$. On the other hand, it means that
we have the boundary-value  but not the Cauchy problem.

\section{Stable quantum oscillator}

Let us remind the canonical quantization of simple oscillator in
quantum mechanics. The Hamiltonian and canonical commutation
relation are
\begin{eqnarray}\label{h1}
&& H={1\over 2}(p^2+\omega^2q^2),~~~~[q,p]=i.
\end{eqnarray}
Let us introduce the variables
\begin{eqnarray*}\label{h2}
&& a=\sqrt{{\omega\over2}}\left(q+{ip\over\omega}\right),~~~~~
a^+=\sqrt{{\omega\over2}}\left(q-{ip\over\omega}\right),~~~~~
[a,a^+]=1.
\end{eqnarray*}
The Hamiltonian becomes the form
\begin{eqnarray*}\label{h1}
&& H={\omega\over 2}\left(a^+a+{1\over2}\right).
\end{eqnarray*}
The following commutation relations take place
\begin{eqnarray*}\label{h2}
&& [H,a]=-\omega a,~~~~~[H,a^+]=\omega a^+.
\end{eqnarray*}
Let the function $\Psi_E$ be a eigenfunction with eigenvalue $E$
\begin{eqnarray*}\label{h3}
&& H\Psi_E=E\Psi_E.
\end{eqnarray*}
One can get
\begin{eqnarray*}\label{h4}
&& Ha\Psi_E=(E-\omega)a\Psi_E,\\
&&Ha^+\Psi_E=(E+\omega)a^+\Psi_E,\\
&& Ha^n\Psi_E=(E-n\omega)a^n\Psi_E,\\
&&H(a^+)^n\Psi_E=(E+n\omega)(a^+)^n\Psi_E.
\end{eqnarray*}
It means that the functions $a^n\Psi_E$ and $(a^+)^n\Psi_E$ are the
eigenfunctions too. Our oscillator is a stable system. It means that
states with arbitrary negative energy should not exist. In order to
guarantee the stability, i.e. positiveness of energy we should
introduce so-called vacuum, or the lowest state $\Psi_0$ which
satisfies
\begin{eqnarray*}\label{h4}
&& (1)~~~~~a\Psi_0=0.
\end{eqnarray*}
In the coordinate representation the vacuum looks like
\begin{eqnarray*}\label{h4}
&& \left(q+{1\over\omega}{d\over
dq}\right)\Psi_0(q)=0,~~~~~\Psi_0(q)=e^{-{\omega\over2}q^2}~\in~{\bf
L}^2.
\end{eqnarray*}
Therefore, vacuum can be normalized
\begin{eqnarray*}\label{h4}
&& (2)~~~~~(\Psi_0^+\Psi_0)=1.
\end{eqnarray*}
Eigenvalue of the vacuum is
\begin{eqnarray*}\label{h4}
&&H\Psi_0=\omega\left[a^+a+{1\over2}\right]\Psi_0={\omega\over2}\Psi_0,~~~E_0={\omega\over2}.
\end{eqnarray*}
Thus, the other eigenfunctions are
\begin{eqnarray*}\label{h4}
&&\Psi_n={(a^+)^n\over
n!}\Psi_0,~~~~~H\Psi_n=\omega\left(n+{1\over2}\right)\Psi_n.
\end{eqnarray*}
The time dependence of these eigenfunction is defined by
\begin{eqnarray*}\label{h4}
&&
\Psi_n(t)=e^{-iHt}\Psi_n=e^{-i\omega\left(n+{1\over2}\right)t}\Psi_n.
\end{eqnarray*}

The macroscopic oscillation is described by the coherence state
$\Psi_f=e^{{f\over\sqrt{2}}a^+}\Psi_0$ where $f$ is the amplitude of
oscillation, so that
\begin{eqnarray*}\label{h4}
&& \langle q(t)\rangle_f={(\Psi_f^+ q(t)\Psi_f)\over(\Psi_f^+
\Psi_f)}={f\over\sqrt{\omega}}\cos(\omega t).
\end{eqnarray*}

The coordinate operator $q(t)$ depends on time as
\begin{eqnarray*}\label{h4}
&& q(t)=e^{iHt}qe^{-iHt}= {1\over\sqrt{2\omega}}\left(ae^{-i\omega
t}+a^+e^{i\omega t}\right).
\end{eqnarray*}
The commutator is
\begin{eqnarray*}\label{h4}
&& \Delta(t-t')=[q(t),q(t')]= {1\over2\omega}\sin\omega(t-t')
\end{eqnarray*}
and causal Green function, or propagator looks as
\begin{eqnarray*}\label{h4}
&& D_c(t-t')=(\Psi_0,{\rm
T}(q(t)q(t'))\Psi_0)={1\over2\omega}e^{-i\omega|t-t'|}.
\end{eqnarray*}
For time ordered exponent we get
\begin{eqnarray*}\label{h4}
&& \left(\Psi_0,{\rm T}\left\{e^{\int\limits_{t_0}^{t_1}
dt~q(t)J(t)}\right\}\Psi_0\right)= e^{{1\over2}\int\!\!\!
\int\limits_{t_0}^{t_1} dt dt'~J(t)D_c(t-t')J(t')}
\end{eqnarray*}

\section{Unstable quantum oscillator}

Now let us apply the stated above procedure of canonical
quantization to unstable oscillator, the Hamiltonian of which is
\begin{eqnarray}\label{h1}
&& H={1\over 2}(p^2-\Omega^2q^2),~~~~[q,p]=i.
\end{eqnarray}
Let us introduce the operators
\begin{eqnarray*}\label{h2}
&& A=\sqrt{{\Omega\over2}}\left(q-{p\over\Omega}\right),~~~~~
B=\sqrt{{\Omega\over2}}\left(q+{p\over\Omega}\right),~~~~[A,B]=i.
\end{eqnarray*}
One should stress the operators $A$ and $B$ are hermitian.

The Hamiltonian becomes of the form
\begin{eqnarray*}\label{h1}
&& H={1\over 2}(p^2-\Omega^2q^2)=\Omega\left(-BA-{i\over2}\right).
\end{eqnarray*}
The following commutation relations take place
\begin{eqnarray*}\label{h2}
&& [H,A]=i\Omega A,~~~~~[H,B]=-i\Omega B.
\end{eqnarray*}
Let the function $\Phi_E$ be an eigenfunction with eigenvalue $E$
\begin{eqnarray*}\label{h3}
&& H\Phi_E=E\Phi_E.
\end{eqnarray*}
One should remark that this equation, being the equation of the
second order, has two independent solutions for any real or complex
$E$.

Then one can get
\begin{eqnarray*}\label{h4}
&& HA\Phi_E=(E+i\Omega)A\Phi_E,\\
&&HB\Phi_E=(E-i\Omega)B\Phi_E,\\
&& HA^n\Phi_E=(E+in\Omega)A^n\Phi_E,\\
&&HB^n\Phi_E=(E-in\Omega)B^n\Phi_E.
\end{eqnarray*}
It means that the functions $A^n\Phi_E$ and $B^n\Phi_E$ are the
eigenfunctions too with complex eigenvalues. The time dependence of
these eigenfunctions is
\begin{eqnarray*}\label{h4}
&&
e^{-iHt}A^n\Phi_E=e^{-i(E+in\Omega)t}A^n\Phi_E=e^{-iEt}e^{+n\Omega t}A^n\Phi_E,\\
&&e^{-iHt}B^n\Phi_E=e^{-i(E-in\Omega)t}B^n\Phi_E=e^{-iEt}e^{-n\Omega
t}B^n\Phi_E.
\end{eqnarray*}
One can see that the eigenfunctions $A^n\Phi_E$ grow  as $e^{nt}$.
Naturally,  it is not physical behavior, so by analogy with
stability requirement we should impose the condition to exclude
growing states. For this aim we introduce a state namad
''pseudo-vacuum'' $\Phi_0$ which satisfies
\begin{eqnarray*}\label{h4}
&& A\Phi_0=0.
\end{eqnarray*}
In the coordinate space it looks like
\begin{eqnarray*}\label{h4}
&& A\Phi_0=\left(q+{i\over \Omega}{d\over
dq}\right)\Phi_0(q)=0,~~~~~\Phi_0(q)=e^{{i\Omega\over2}q^2}\in\!\!\!\!\!/~{\bf
L}^2.
\end{eqnarray*}
"Vacuum"$~$energy is equal to
\begin{eqnarray*}\label{h4}
&&H\Phi_0=\Omega\left[-BA-{i\over2}\right]\Phi_0=-{i\Omega\over2}\Phi_0,~~~E_0=-{i\Omega\over2}.
\end{eqnarray*}
The eigenfunctions and eigenvalues are
\begin{eqnarray*}\label{h4}
&&\Phi_n={B^n\over\sqrt{n!}}\Phi_0,\\
&&H\Phi_n=-iE_n\Phi_n,~~~~~~~~E_n=\Omega\left(n+{1\over2}\right).
\end{eqnarray*}
The time dependence of these states is
\begin{eqnarray*}\label{h4}
&&\Phi_n(t)=e^{-iHt}\Phi_n=e^{-\Omega\left(n+{1\over2}\right)t}\Phi_n.
\end{eqnarray*}
It means that all states disappear when time increases $t\to\infty$.

Thus, the unstable hamiltonian has a pure imaginary spectrum. It
means that the standard methods of functional analysis of hermitian
operators in the functional space ${\bf L}^2$ are not applicable in
our case. This situation is similar to problems with indefinite
metric (see \cite{Nag} and Appendix). We should introduce the rules
how to calculate the matrix elements of operators depending on $A$
and $B$. One can write
\begin{eqnarray*}\label{h4}
&& (\Phi_1,F(B)\Phi_2)=\int dq~ \Phi_1(q)\cdot F\left(q+{1\over
i\Omega}{d\over dq}\right)\Phi_2(q)\\
&&=\int dq~ F\left(q-{1\over i\Omega}{d\over
dq}\right)\Phi_1(q)\cdot \Phi_2(q)=(F(A)\Phi_1,\Phi_2).
\end{eqnarray*}

The macroscopic motion is described by the coherence state
\begin{eqnarray*}\label{h4}
&&
\Phi_f=e^{-i{f\over\sqrt{2}}B}\Phi_0,~~~~~~~~\Phi_f(t)=e^{-iHt}\phi_f=e^{-{\Omega\over2}
t}e^{-i{f\over\sqrt{2}}Be^{-\Omega t}}\Phi_0,
\end{eqnarray*}
and for the  coordinate $q={A+B\over\sqrt{2\Omega}}$ one can get
\begin{eqnarray*}\label{h4}
&& \langle
q(t)\rangle_f={(\Phi_f(t),q\Phi_f(t))\over(\Phi_f(t),\Phi_f(t))}
={\left(e^{-i{f\over\sqrt{2}}Be^{-\Omega
t}}\Phi_0,{A+B\over\sqrt{2\Omega}}e^{-i{f\over\sqrt{2}}Be^{-\Omega
t}}\Phi_0\right)\over
\left(e^{-i{f\over\sqrt{2}}Be^{-\Omega t}}\Phi_0,e^{-i{f\over\sqrt{2}}Be^{-\Omega t}}\Phi_0\right)}\\
&&=\left.{\left(\Phi_0,e^{-i{f_1\over\sqrt{2}}Ae^{-\Omega
t}}{A+B\over\sqrt{2\Omega}}e^{-i{f_2\over\sqrt{2}}Be^{-\Omega
t}}\Phi_0\right)\over
\left(\Phi_0,e^{-i{f_1\over\sqrt{2}}fAe^{-\Omega t}}e^{-i{f_2\over\sqrt{2}}Be^{-\Omega t}}\Phi_0\right)}\right|_{f_1=f_2=f}\\
&&=\left.i{e^{\Omega
t}\over\sqrt{2\Omega}}\left({\partial\over\partial
f_1}+{\partial\over\partial f_2}\right)
\ln\left(\Phi_0,e^{-i{f_1\over\sqrt{2}}Ae^{-\Omega t}}\cdot
e^{-i{f_2\over\sqrt{2}}Be^{-\Omega
t}}\Phi_0\right)\right|_{f_1=f_2=f}.
\end{eqnarray*}
The last matrix element can be calculated as
\begin{eqnarray*}\label{h4}
&&\left(\Phi_0,e^{-i{f_1\over\sqrt{2}}Ae^{-\Omega t}}\cdot
e^{-i{f_2\over\sqrt{2}}Be^{-\Omega
t}}\Phi_0\right)=\left(\Phi_0,e^{-i{f_1\over\sqrt{2}}Be^{-\Omega
t}}\cdot e^{-{f_1f_2\over2}e^{-2\Omega t}[A,B]}\cdot
e^{-i{f_2\over\sqrt{2}}Ae^{-\Omega
t}}\Phi_0\right)\\
&&=e^{-i{f_1f_2\over2}e^{-2\Omega t}}\left(\Phi_0,\Phi_0\right).
\end{eqnarray*}

As a result the unstable oscillator decays exponentially
\begin{eqnarray*}\label{h4}
&& \langle q(t)\rangle_f={f\over\sqrt{2\Omega}}e^{-\Omega t}.
\end{eqnarray*}

Let us write down the useful formulas. The coordinate operator is
\begin{eqnarray*}\label{h4}
&& q(t)=e^{iHt}qe^{-iHt}=e^{iHt}{A+B\over\sqrt{2\Omega}}e^{-iHt}=
{1\over\sqrt{2\Omega}}\left(Ae^{-\Omega t}+Be^{\Omega t}\right).
\end{eqnarray*}
The commutator is written as
\begin{eqnarray*}\label{h4}
&& [q(t),q(t')]= {1\over2\Omega}\left[Ae^{-\Omega t}+Be^{\Omega
t},Ae^{-\Omega t'}+Be^{\Omega
t'}\right]=-i~{\sinh\Omega(t-t')\over\Omega}
\end{eqnarray*}
and the causal Green function, or propagator looks like
\begin{eqnarray*}\label{h4}
&& D_c(t-t')=\langle{\rm T}(q(t)q(t'))\rangle_0= {1\over2\Omega}
\left\langle{\rm
T}\left((Ae^{-\Omega t}+Be^{\Omega t})(Ae^{-\Omega t'}+Be^{\Omega t'})\right)\right\rangle_0\\
&&={i\over2\Omega}e^{-\Omega|t-t'|}.
\end{eqnarray*}
We would like to stress that the factor $''i''$ has sense of the
''norm'' of the unstable state. Here we see an analogy with the
''norm'' of indefinite states, where this ''norm'' is equal to $-1$.

The $T-$exponent is equal to
\begin{eqnarray*}\label{h4}
&& \left\langle{\rm T}\left\{e^{\int\limits_{t_0}^{t_1}
dt~q(t)J(t)}\right\}\right\rangle_0= e^{{1\over2}\int\!\!\!
\int\limits_{t_0}^{t_1} dt dt'~J(t)D_c(t-t')J(t')}
\end{eqnarray*}

\newpage

\section{Tachyon field in  QFT}

The Lagrangian of the scalar tachyon field looks like
\begin{eqnarray*}\label{h4}
&&L(t)={1\over2}\int d{\bf x} \left[(\dot{\phi}(t,{\bf
x}))^2-(\nabla\phi(t,{\bf x}))^2+m^2\phi^2(t,{\bf x})\right].
\end{eqnarray*}
Let us introduce the canonical momenta $\pi(t,{\bf x})={\delta
L(t)\over\delta\dot{\phi}(t,{\bf x})}=\dot{\phi}(t,{\bf x})$ and
write down the Hamiltonian
\begin{eqnarray*}\label{h4}
&&H={1\over2}\int d{\bf x} \left[(\pi({\bf x}))^2+(\nabla\phi({\bf
x}))^2-m^2\phi^2({\bf x})\right].
\end{eqnarray*}
The canonical commutation relations are
\begin{eqnarray*}\label{h4}
&&[\phi({\bf x}),\pi({\bf x}')]=i\delta({\bf x}-{\bf x}').
\end{eqnarray*}
It is convenient to go to the momentum representation
$$\phi({\bf x})=\int {d{\bf k}\over(2\pi)^{{3\over2}}}\phi({\bf k})e^{i({\bf kx})},~~~~
\pi({\bf x})=\int {d{\bf k}\over(2\pi)^{{3\over2}}}\pi({\bf
k})e^{-i({\bf kx})}$$
with the canonical commutation relation
\begin{eqnarray*}\label{h4}
&&[\phi({\bf k}),\pi({\bf k}')]=i\delta({\bf k}-{\bf k}').
\end{eqnarray*}
The Hamiltonian looks like ($k^2={\bf k}^2$)
\begin{eqnarray*}\label{h4}
&& H={1\over2}\int d{\bf k}\left[\pi^2({\bf k})+(k^2-m^2)\phi^2({\bf
k})\right]=H_{in}+H_{st}
\end{eqnarray*}
where
\begin{eqnarray*}\label{h4}
&& H_{st}={1\over2}\int d{\bf k}~\theta(k^2-m^2)\left[\pi^2({\bf
k})+(k^2-m^2)\phi^2({\bf k})\right],\\
&& H_{un}={1\over2}\int d{\bf k}~\theta(m^2-k^2)\left[\pi^2({\bf
k})-(m^2-k^2)\phi^2({\bf k})\right].
\end{eqnarray*}
Both Hamiltonian describe a set  of independent oscillators, but the
hamiltonian $H_{st}$ with $k^2>m^2$ contains the real energies
$E_k=\pm\sqrt{k^2-m^2}$ and corresponds to standard stable picture.
However, the Hamiltonian $H_{un}$ with $k^2<m^2$ contains the pure
imaginary energies $E_k=\pm i\sqrt{m^2-k^2}$ and corresponds to the
instable picture.

The quantization of these two hamiltonian can be performed according
procedure described in the previous sections..

\newpage

\subsection{Stable region ${\bf k^2>m^2}$}

In the stable case the standard quantization procedure can be
applied. Let us introduce the operators
\begin{eqnarray*}\label{h4}
&& a_{\bf k}=\sqrt{{\omega_k\over2}}\left[\phi({\bf k})+i{\pi({\bf
k})\over\omega_k}\right],~~~~
a_{\bf k}^+=\sqrt{{\omega_k\over2}}\left[\phi({\bf k})-i{\pi({\bf k})\over\omega_k}\right],\\
&&[a_{\bf k},a_{{\bf k}'}^+]=\delta({\bf k}-{\bf k}')
\end{eqnarray*}
where $\omega_k=\sqrt{k^2-m^2}$. One can see that
\begin{eqnarray*}\label{h4}
&& [H_{st},a_{\bf k}]=-\omega_ka_{\bf k},~~~~[H_{st},a_{\bf
k}^+]=\omega_ka_{\bf k}^+.
\end{eqnarray*}
The lowest state, or vacuum is defined by the condition
\begin{eqnarray*}\label{h4}
&& a_{\bf k}\Psi_0=0.
\end{eqnarray*}
Then the stable part of the tachyon field has the standard form
\begin{eqnarray*}\label{h4}
&& \eta_{\bf k}={1\over\sqrt{2\omega_k}}(a_{\bf k}+a_{\bf k}^+),\\
&&\eta_{\bf k}(t)=e^{iH_{st}t}\phi_{\bf k}
e^{-iH_{st}t}={1\over\sqrt{2\omega_k}}\left(a_{\bf k}~e^{-i\omega_k
t}+a_{\bf k}^+~e^{i\omega_k t}\right),\\
&&\eta(t,{\bf x})=\int{d{\bf
k}\over(2\pi)^{{3\over2}}}{\theta(k^2-m^2)\over\sqrt{2\omega_k}}\left(a_{\bf
k}~e^{-i\omega_kt+i{\bf kx}} +a_{\bf k}^+~e^{i\omega_k t-i{\bf
kx}}\right)
\end{eqnarray*}

\subsection{Unstable region ${\bf k^2<m^2}$}

In the unstable case the  quantization procedure developed for
unstable oscillator can be applied. Let us introduce the operators
\begin{eqnarray*}\label{h4}
&& A_{\bf k}=\sqrt{{\Omega_k\over2}}\left[\phi({\bf k})-{\pi({\bf
k})\over\Omega_k}\right],~~~ B_{\bf
k}=\sqrt{{\Omega_k\over2}}\left[\phi({\bf k})+{\pi({\bf
k})\over\Omega_k}\right],
\end{eqnarray*}
where $\Omega_k=\sqrt{m^2-k^2}$. The commutation relations are
\begin{eqnarray*}\label{h4}
&&[A_{\bf k},B_{{\bf k}'}]=i\delta({\bf k}-{\bf k}').
\end{eqnarray*}
These commutation relations lead to
\begin{eqnarray*}\label{h4}
&& [H_{in},A_{\bf k}]=i\Omega_kA_{\bf k},~~~~[H_{in},B_{\bf
k}]=-i\Omega_kB_{\bf k}.
\end{eqnarray*}
Let the function $\Phi_E$ be an eigenfunction with eigenvalue $E$
\begin{eqnarray*}\label{h3}
&& H\Phi_E=E\Phi_E.
\end{eqnarray*}
Then one can get
\begin{eqnarray*}\label{h4}
&& HA_{\bf k}\Phi_E=(E+in\Omega_k)A_{\bf k}\Phi_E,\\
&&HB_{\bf k}\Phi_E=(E-in\Omega_k)B_{\bf k}\Phi_E.
\end{eqnarray*}
It means that the functions $A_{\bf k}\Phi_E$ and $B_{\bf k}\Phi_E$
are the eigenfunctions too with complex eigenvalues. The time
dependence of these eigenfunctions is
\begin{eqnarray*}\label{h4}
&& e^{-iHt}A_{\bf k}\Phi_E=e^{-iEt+\Omega_k t}A_{\bf k}\Phi_E,\\
&& e^{-iHt}B_{\bf k}\Phi_E=e^{-iEt-\Omega_k t}A_{\bf k}\Phi_E.
\end{eqnarray*}
One can see that the eigenfunctions $A_{\bf k}\Phi_E$ grow  as
$e^{\Omega t}$. From a standard point of view it is not physical
behavior, so by analogy with the stability requirement we should
impose the condition, namely, we introduce a state - pseudo-vacuum
$\Phi_0$ which satisfies
\begin{eqnarray*}\label{h4}
&& A_{\bf k}\Phi_0=0.
\end{eqnarray*}
All further argumentations and calculations repeat literally the
results of the section 4. As a result we have for the instable part
of the tachyon field
\begin{eqnarray*}\label{h4}
&& \chi_{\bf k}={1\over\sqrt{2\Omega_k}}(A_{\bf k}+B_{\bf k}),\\
&&\chi_{\bf k}(t)=e^{iH_{in}t}\phi_{\bf k}
e^{-iH_{in}t}={1\over\sqrt{2\Omega_k}}\left(A_{\bf k}~e^{-\Omega_k
t}+B_{\bf k}~e^{\Omega_k t}\right),\\
&&\chi(t,{\bf x})=\int{d{\bf
k}\over(2\pi)^{{3\over2}}}{\theta(m^2-k^2)\over\sqrt{2\Omega_k}}\left(A_{\bf
k}~e^{-\Omega_kt+i{\bf kx}} +B_{\bf k}~e^{\Omega_k t-i{\bf
kx}}\right).
\end{eqnarray*}

\subsection{The tachyon field}

The tachyon field is a sum of stable and unstable fields
\begin{eqnarray*}\label{h4}
&& \phi(t,{\bf x})=\eta(t,{\bf x})+\chi(t,{\bf x})
\end{eqnarray*}
for which the vacuum is
\begin{eqnarray*}\label{h4}
&& |0\rangle=\Psi_0\Phi_0.
\end{eqnarray*}

One tachyon states are
\begin{eqnarray*}\label{h4}
&& \langle0|\phi(t,{\bf x})a_{\bf
k}^+|0\rangle={1\over(2\pi)^{{3\over2}}\sqrt{2\omega_k}}~e^{-i\omega_kt+i{\bf
kx}}\\
&& \langle0|\phi(t,{\bf x})B_{\bf
k}|0\rangle={i\over(2\pi)^{{3\over2}}\sqrt{2\Omega_k}}~e^{-\Omega_kt+i{\bf
kx}}
\end{eqnarray*}
One can see that the unstable states decrease exponentially.

The causal Green function, or propagator looks like
\begin{eqnarray*}\label{h4}
&& D_c(t-t',{\bf x-x'})=\langle T(\phi(t,{\bf x})\phi(t',{\bf x}'))\rangle_0\\
&&=\int{d{\bf k}\over(2\pi)^3}{\theta(k^2-m^2)\over2\sqrt{k^2-m^2}}
e^{-i\omega_k|t-t'|+i{\bf k(x-x')}}+i\int{d{\bf
k}\over(2\pi)^3}{\theta(m^2-k^2)\over2\sqrt{m^2-k^2}}
e^{-\Omega_k|t-t'|+i{\bf k(x-x')}}\\
\end{eqnarray*}
This representation can be represented in the usual form
\begin{eqnarray*}\label{h4}
&& D_c(t,{\bf x})=\int\limits_m^\infty{dk~k^2\over(2\pi)^2
2\sqrt{k^2-m^2}} e^{-i\omega_k|t-t'|}{\sin(kx)\over xk}
+i\int\limits_0^m{dk~k^2\over(2\pi)^2 2\sqrt{m^2-k^2}}
e^{-\Omega_k|t-t'|}{\sin(kx)\over xk}\\
&&=\int\limits_C{dk~k^2\over(2\pi)^2 2\sqrt{k^2-m^2}}
e^{-i\omega_k|t-t'|}{\sin(kx)\over xk}\\
&&=\int{d^4q\over(2\pi)^4i}\cdot{e^{-iqx}\over
(q^2+m^2+i0)},~~~q^2=q_0^2-{\bf q}^2.
\end{eqnarray*}

\section{Conclusion}

The asymptotic tachyon field contains only the stable field
\begin{eqnarray*}\label{h4}
&& \phi_{in}(t,{\bf x})=\phi_{out}(t,{\bf x})=\eta(t,{\bf x})
\end{eqnarray*}
for which the vacuum is the standard Fock vacuum $|0\rangle=\Psi_0$
with one particle state
\begin{eqnarray*}\label{h4}
&& \langle0|\eta(t,{\bf x})a_{\bf
k}^+|0\rangle={1\over(2\pi)^{{3\over2}}\sqrt{2\omega_k}}~e^{-i\omega_kt+i{\bf
kx}},~~~~~~\omega_k^2=k^2-m^2>0.
\end{eqnarray*}
It means that the unstable tachyon components do not exist as
physical states.

The formula
\begin{eqnarray*}\label{h4}
&& \left\langle0\left|{\rm T}\left\{e^{\int dx
~\phi(x)J(x)}\right\}\right|0\right\rangle= e^{{1\over2}\int\!\!\!
\int dxdx'~J(x)D_c(x-x')J(x')}
\end{eqnarray*}
with
\begin{eqnarray*}\label{h4}
&& D_c(t,{\bf x})=\int{d^4q\over(2\pi)^4i}\cdot{e^{-iqx}\over
(q^2+m^2+i0)},~~~q^2=q_0^2-{\bf q}^2
\end{eqnarray*}
permits one to construct the $S$-matrix in the form of the standard
perturbation decomposition in the form of Feynman diagrams.

Our scheme is similar to quantization of the electro-magnetic field
when the photon field has two physical (transverse) and two
nonphysical (longitudinal and time) components. The nonphysical
components do not exist as physical states, the norm of these states
is not a positive number, but they are needed to construct the
photon causal propagator.

The next step is to introduce an interaction of the tachyon with
other particles, calculate amplitudes for different processes and
analyze these amplitudes of perturbation expansion, but it is a
different story with numerous problems.

\section{Appendix}

Let us introduce the linear space with complex metric ${\cal H}_c$.
A vector $|\Phi\rangle$ belongs to ${\cal H}_c$, if
\begin{enumerate}
\item The scalar product $(\Phi_1,\Phi_2)$ is defined;
\item The scalar product $(\Phi,\Phi)$ is a complex number;
\item There exist two operators $A$ and $B$ which satisfy the conditions
\begin{itemize}
\item if $\Phi\in{\cal H}_c$, then $A\Phi\in{\cal H}_c$ and $B\Phi\in{\cal
H}_c$~;
\item commutation relations are
\begin{eqnarray}\label{AI1}
&&[A,B]=i;
\end{eqnarray}
\item the following equalities take place:
\begin{eqnarray}\label{AI2}
&&(\Phi_1,B\Phi_2)=(A\Phi_1,\Phi_2),~~~(\Phi_1,A\Phi_2)=(B\Phi_1,\Phi_2);
\end{eqnarray}
\end{itemize}
\item There exists the vector, named ''pseudo-vacuum'', $\Phi_0\in{\cal H}_c$,
which satisfies
\begin{eqnarray}\label{AI3}
&&(1)~~~A\Phi_0=0,~~~~~(2)~~~ (\Phi_0,\Phi_0)\neq0.
\end{eqnarray}
\end{enumerate}

The vector space ${\cal H}_c$ occupies an intermediate place between
the standard Hilbert space and the vector space with indefinite
metric (see \cite{Nag}).

If the conditions (\ref{AI1}),(\ref{AI2}) and (\ref{AI3}) take
place, then one can get
\begin{eqnarray}\label{AI4}
&& \left(e^{f_1B}\Phi_0, e^{f_2B}\Phi_0\right)=
\left(\Phi_0,e^{f_1A}\cdot e^{f_2B}\Phi_0\right)
=\left(\Phi_0,e^{f_2B}\cdot e^{f_1f_2[A,B]}\cdot
e^{f_1A}\Phi_0\right)\nonumber\\
&&=\left(e^{f_2A}\Phi_0, e^{if_1f_2}\cdot
e^{f_1A}\Phi_0\right)=e^{if_1f_2}\left(\Phi_0,\Phi_0\right)
\end{eqnarray}
and
\begin{eqnarray}\label{AI5}
&& M_{f_1,f_2}={\left(e^{f_1B}\Phi_0, e^{f_2B}\Phi_0\right)\over
\left(\Phi_0,\Phi_0\right)}=e^{if_1f_2}.
\end{eqnarray}

\end{document}